\documentclass[
    ,final            
  ]
  {aipproc}

\layoutstyle{8x11single}

\usepackage{amssymb,latexsym}
\usepackage{amsmath}

\newcommand{\gsm}{G_{\rm SM}}
\newcommand{\ggut}{G_{\rm GUT}}
\newcommand{\gps}{G_{\rm PS}} 
\newcommand{\mgut}{M_{\rm GUT}}
\newcommand{\Mpl}{M_{\rm Pl}} 

\newcommand{\bl}{U(1)_{\rm B-L}}

\begin{document}

\title{Constraining SUSY GUTs and Inflation with Cosmology}


\author{Jonathan Rocher\footnote{Email: Jonathan.Rocher@utsa.edu. My current address is at the University of Texas.}}{
  address={Institut d'Astrophysique de Paris, 98bis boulevard Arago, 75014 Paris, France, \\ and, \\}
,altaddress={Department of Physics and Astronomy, University of Texas - San Antonio, \\
6900 N. loop 1604 W., San Antonio, TX 78249, USA.}
}

\begin{abstract}
In the framework of Supersymmetric Grand Unified Theories (SUSY 
GUTs), the
universe undergoes a cascade of symmetry breakings, during which
topological defects can be formed. We address the question of the
probability of cosmic string formation after a phase of hybrid
inflation within a large number of models of SUSY GUTs in
agreement with particle and cosmological data. We show that cosmic
strings are extremely generic and should be used to relate
cosmology and high energy physics. This conclusion is employed
together with the WMAP CMB data to strongly constrain SUSY hybrid
inflation models. F-term and D-term inflation are studied in the
SUSY and minimal SUGRA framework. They are both found to agree
with data but suffer from fine tuning of their superpotential
coupling ($\lambda \lesssim 3\times 10^{-5}$ or less). Mass scales
of inflation are also constrained to be less than $M \lesssim
3\times 10^{15}$ GeV. 
\end{abstract}

\maketitle


\section{Introduction}
The Supersymmetric Grand Unified Theories (SUSY GUTs) are a
motivated framework in which to describe the early universe physics
\cite{Langacker:physrept,Mohapatra:review,Mohapatra:book}. The
symmetries of the Standard Model (SM) $\gsm\equiv SU(3)_C\times
SU(2)_L\times U(1)_Y$ are then supposed to be obtained at low
energies from a larger gauge group $\ggut$ through a cascade of
Spontaneous Symmetry Breaking (SSB). According to the Kibble
mechanism~\cite{kibble:76,vilshel}, topological defects should
form during the phase transition associated with these SSBs.
However, it is well known that the formation of domain walls or
monopoles would rapidly dominate the dynamics of the universe, in
contradiction with observations. The situation for Cosmic
Strings (CS) however is very different since a network of such
objects reaches a phase of scaling where their relative energy stays
constant with time. With the current observational precision 
however, they have not yet been detected, for example in the WMAP 
CMB data.

In this review, we will first quantify how probable
is the cosmic string formation within the SUSY GUT framework in a
large number of motivated and realistic models. This class of models
is chosen to be in agreement with proton decay data, observed
neutrino oscillations, baryon asymmetry and CMB anisotropies. Any
realistic model must include a phase of inflation and we need to
know whether the formation of strings occurs before or after the last
inflation phase. Indeed, in the last case only, they can have an
observable effect on cosmology. This analysis allows us to conclude
that cosmic string are a highly generic prediction of SUSY GUTs.

In a second section, we confront this prediction with the WMAP CMB
data (first year release). The low allowed contribution from
strings to CMB anisotropies is employed to constrain strongly the
parameters of the two motivated inflationary models : F-term and
D-term SUSY hybrid inflation. We will show that all parameters can
be constrained, including superpotential coupling constants and
mass scale of inflation. This allows us to re-analyse the naturalness
of these models.

\section{Genericity of cosmic string formation in SUSY GUTs}
\subsection{Model building}
A model for the early universe, to be realistic, should explain
\cite{prd2003,theseJO} the observed oscillation of neutrinos, be
in agreement with lower bounds on proton decay, explain the
baryon/anti-baryon asymmetry in the universe, and be consistent 
with the most recent
CMB data. SUSY GUTs offer a coherent framework for constructing
such models, the unification being suggested by the running of SM
coupling constants and allowing one to reduce the number of free
parameters of the SM. Supersymmetry is motivated since it
offers an elegant solution to the hierarchy problem and it allows
a unification at sufficiently high energy ($\mgut \sim 2\times
10^{16}$ GeV) for the dimension 6 operators of proton decay to be
suppressed. Within SUSY GUTs, the most economical way to protect
the proton from operators of dimension 4 (exchange of sfermions)
is by keeping unbroken at low energy the R-parity. This R-parity
is a $Z_2$ symmetry defined by \cite{Rparity}
$$R=(-1)^{2S+3(B-L)}~,$$ where $S$ is the spin, $B$ is the baryon
number and $L$ is the lepton number of a particle. We can easily
check that if this parity is conserved, the decay of the Lightest
Supersymmetric Particle (LSP) is forbidden, introducing a good
candidate for dark matter without any new assumption. This
R-parity can be preserved by breaking a GUT group containing the
$\bl$ symmetry with an even-charged representation \cite{spmartin}.
It will be necessary to identify how this abelian symmetry is
embedded in $\ggut$ to make sure no SSB patterns incompatible with
proton decay are considered, i.e. R-parity is kept unbroken. For
model builders, this also means that the choice of Higgs
representations employed has to be done carefully.

The observed neutrino oscillations \cite{SuperK98} require that
these particles are massive and thus that we go beyond the SM. We
then need to introduce a right-handed (RH) neutrino, singlet under
the SM gauge group. The most economical way to generate small mass
neutrinos is via a see-saw mechanism, ie generate a large Majorana
mass $M_N$ of the GUT scale and a Dirac mass of the electroweak
scale $M_\nu$. The mass eigenstates are then of the order of $M_N$
and $M_\nu^2/M_N$ which allows one to explain current limits on light
neutrino masses. The Majorana mass term is usually generated
through the breaking of the $\bl$ symmetry within GUTs. This will
induce some constraints on the GUT groups considered in our study
since we will assume that they contain the $\bl$ symmetry.

The current CMB data clearly favor an inflationary origin for
generation of the primordial anisotropies (even if a subdominant
contribution from cosmic strings is not excluded, see next
section). A phase of inflation in the early universe history is 
thus almost necessary to explain these data, or to solve the 
monopole problem and other problems of the standard model of
cosmology. Within our framework, the most natural model of
inflation is the one of hybrid inflation. It is based on the
superpotential \cite{dvashascha}
\begin{equation}\label{ftermsuperpotential}
W_{\rm F}=\kappa S (\Phi_+ \Phi_- - M^2)~.
\end{equation}
It couples an inflaton field $S$ to a pair of GUT Higgs fields
$\Phi_+$ and $\Phi_-$ non trivially charged under a symmetry $G$.
The model possesses only two parameters : a coupling constant
$\kappa$ and a mass scale $M$. The fields $\Phi_\pm$ must belong
to non trivial complex conjugate representations of $G$ and the
dynamics of the system induce a Vacuum Expectation Value (VEV) for
these fields, which breaks the symmetry $G$. This induces a
waterfall that ends the inflationary phase. In F-term hybrid
inflation, this symmetry $G$ can be part of $\ggut$ but in the
case of D-term inflation, $G=U(1)$ and this abelian factor cannot
be a subgroup of a non-abelian group, since the model requires a 
non-vanishing Fayet-Iliopoulos term $\xi$. In this section we will
assume a GUT model based on a simple group $\ggut$ to unify the
coupling constant at high energy and thus restrict ourselves to
F-term inflation. This model is motivated in our framework since
\begin{itemize}
\item no new fields have to be introduced (except the inflaton
singlet) since GUT Higgs fields are already present. Note that
this singlet can also be a Higgs singlet already present in the
theory.

\item the form of Eq~(\ref{ftermsuperpotential}) is the most
general that is renormalizable, invariant under $G$ and under an
R-symmetry.

\item the potential is naturally flat (even perfectly at tree
level) and the radiative corrections don't spoil the model but
introduce a lift in the potential that helps the slow roll.

\item F-term inflation is known to become unstable under
Supergravity (SUGRA) corrections; but in this model, the dangerous
terms happen to cancel in a minimal SUGRA. Moreover, the value of
the inflaton field in the observable range stays far below the
Planck scale \cite{jcap2005}. This shows that the SUGRA
corrections may be safely neglected.

\item the model has been introduced to explain the level of CMB
anisotropies with a non fine tuned coupling $\kappa$ ($10^{-2}$).
This major motivation will however be re-analyzed in the next
section.
\end{itemize}
The consequences of this choice of inflation model on our study
are important : inflation is then assumed to occur before one of
the SSB of GUT patterns. We remind the reader that this inflation
has to occur after the last formation of monopoles to avoid
contradiction with cosmology.

The baryon/anti-baryon asymmetry can naturally be explained through a first
stage of leptogenesis in our framework. This model is able to
explain the observed level of asymmetry and does not require any new
ingredient, if we have already introduced RH neutrinos for
see-saw. The lepton asymmetry is then generated by the decay of RH
neutrinos. The baryon asymmetry is generated subsequently through
sphaleron transitions. Since a phase of inflation would wash out
any existing asymmetry, this phase has to occur after the end of
inflation. The standard model of leptogenesis is called thermal,
which means that the decay of RH neutrinos occurs when the
temperature of the universe drops below their mass. However, RH
neutrino masses are constrained by the light neutrino masses and
the reheating temperature is constrained by nucleosynthesis and
gravitino production. This induces a tension in the parameter
space for the thermal model to take place. One way to get out is
to consider a small extension of the model : non thermal
leptogenesis. We will assume it in this study since it requires
two ingredients we already assumed : an F-term hybrid inflation
and the see-saw mechanism. This model assumes that the
right-handed neutrinos get a mass through a coupling with one of
the $\Phi_\pm$ Higgs fields. At the end of inflation, the inflaton
and Higgs fields start to oscillate and disintegrate into
right-handed neutrinos $S\rightarrow N_R N_R$. Then, for this to
be allowed, we just need the mass of the inflaton field to be
higher that twice the mass of one right-handed neutrino and no
tension exists on the reheating temperature. The consequence of
this assumption on our study is to impose the see-saw mechanism
and thus the breaking of $\bl$ (or a group that contains it) after
inflation.

\subsection{GUT gauge groups and SSB patterns}
We consider in this study only simple GUT gauge groups. There are
several constraints on them to be eligible (minimal rank of 4,
complex and anomaly free fermionic representations). Only
simple groups of rank lower than 8 are considered : $SU(n)$, with
$n\leq 9$, $SO(10)$, $SO(14)$, and $E_6$. Larger groups would
mainly generate a lot of similar or identical SSB patterns, and
lose a lot of predictability. Moreover, if we require that $ \bl
\subset \ggut$, then $n\geq 7$ .

To write down all possible SSB pattern from $\ggut$ down to $\gsm
\times Z_2$, we must know how the SM gauge group $\gsm$ is
embedded inside $\ggut$ in order not to break part of $\gsm$. We
also need to know the possible embeddings of $\bl$ in order to
know when the R-parity can appear. For example, if we consider
$SO(10)$ broken through the Pati-Salam group $\gps\equiv
SU(4)_C\times SU(2)_L\times SU(2)_R$, the SM can be embedded
following
\begin{itemize}
\item $SU(4)_C\supset SU(3)_C \times \bl$~,

\item $Y/2= I_3^R \pm (B-L)/2$~,
\end{itemize}
where $Y$, $(B-L)$ and $I_3^R$ denote respectively the generator
of hypercharge, of $\bl$ and the diagonal generator of $SU(2)_R$.
As a consequence, the only possible SSB schemes from $SO(10)$
through $\gps$ are \cite{prd2003}
\begin{equation}\label{schemasso10}
\begin{array}{ccclllcccc}
 SO(10) & \overset{0}{\rightarrow} & 4_{\rm C}     ~2_{\rm L}     ~2_{\rm R}   &
\left\{
\begin{array}{cllllccc}
\stackrel{1}{\longrightarrow} & 3_{\rm C} ~2_{\rm L} ~2_{\rm R}
~1_{\rm B-L} & \left\{
\begin{array}{cllllccc}
 \stackrel{1}{\longrightarrow}  &   3_{\rm C}     ~2_{\rm L}     ~1_{\rm R}     ~1_{\rm B-L}   &  \stackrel{2}{\longrightarrow}  &   {\rm G}_{\rm SM}  ~Z_2   \\
  \stackrel{2',2}{\longrightarrow}  &   {\rm G}_{\rm SM} ~Z_2\\
 \end{array}
\right.
\\
  \stackrel{1}{\longrightarrow}  &   4_{\rm C}     ~2_{\rm L}     ~1_{\rm R}   &
\left\{
\begin{array}{cllllccc}
  \stackrel{1}{\longrightarrow}  &   3_{\rm C}     ~2_{\rm L}     ~1_{\rm R}     ~1_{\rm B-L}   &   \stackrel{2}{\longrightarrow}  &   {\rm G}_{\rm SM} ~Z_2\\
 \stackrel{2',2}{\longrightarrow}  &   {\rm G}_{\rm SM} ~Z_2\\
 \end{array}
\right.
\\
  \stackrel{1}{\longrightarrow}  &   3_{\rm C}     ~2_{\rm L}     ~1_{\rm R}     ~1_{\rm B-L}   &  ~~~~\stackrel{2}{\longrightarrow}      ~~{\rm G}_{\rm SM} ~Z_2\\
  \end{array}
\right.
\end{array}
\end{equation}
where all notations are the same as in \cite{prd2003}. These
schemes must be read from left to right, each arrow being a SSB
and each line being a different model. The number over the arrows
gives the nature of topological defects (if any) formed during the
SSB : $0$ for no topological defects, $1$ for monopoles, $2$ for
cosmic strings, and $2'$ for embedded strings. We remind the
reader that homotopy group analysis permit us to study the formation
of topological defect \cite{vilshel}. For a given SSB $G
\rightarrow H$, the vacuum manifold of interest is $M=G/H$. The
topological properties of $M$ control the formation of topological
defects : $\pi_0[M]\neq I$ is the condition for domain wall
formation, $\pi_1[M]\neq I$ is the condition for cosmic string
formation and $\pi_2[M]\neq I$ is the condition for monopole
formation. The condition for embedded string formation
\cite{embedded} is similar but apply on $M_{\rm emb}=G_{\rm
emb}/H_{\rm emb}$ where $G_{\rm emb}\subset G$ and $H_{\rm emb}
\subset H$. There are additional conditions for their formation,
concerning the properties of the defect solution.

The other possible schemes for $SO(10)$ are obtained by breaking
$SO(10)$ down to each of the groups that appear in
Eq.~(\ref{schemasso10}), since we want to consider minimal schemes
(only one group is broken at each step) and more direct schemes.\\

We can see on Eq.~(\ref{schemasso10}) that if one now wants to
couple the hybrid inflation phase to one of the SSB for each
model, only one choice is possible. It must occur at the last SSB,
and we can read that this SSB that ends the inflation always
produces cosmic strings. This was to illustrate the result of
\cite{prd2003} where all groups of rank less than 8 are considered
and all known embeddings of the SM have been used to write every
possible SSB pattern. Based on $SO(10)$, we find 34 viable models
that keep R-parity unbroken at low energy and all of them produce
cosmic strings at the end of inflation. Because of the smallness
of this group, the mass per unit length of the strings is directly
related to the inflation energy scale. When considering larger GUT
groups, the number of possibilities increases dramatically but the
conclusion remains identical. The formation of topological cosmic
strings is unavoidable within our assumptions. If we relax the
assumption of thermal leptogenesis, then $98\%$ of $E_6$ based
models produce topological strings. We then conclude
\cite{prd2003} that cosmic strings are \emph{generic} and that
their mass scale is generically proportional to the energy scale
of hybrid inflation. We study in the next section the implication
of this result for hybrid inflationary models.

\section{Constraints on inflationary physics}
This second part is motivated by the results of the previous
section. We wish to study the consequences of cosmic string
formation, generically at the end of the hybrid inflation phase.
We will first consider the F-term SUSY hybrid model and study the
constraints coming from cosmology on its parameters, namely the
coupling constant and the energy scale of inflation. These
constraints will come from the fact that a highly subdominant 
contribution of cosmic strings to the CMB anisotropies is imposed 
by the current CMB data. We will also consider the D-term
SUSY/SUGRA hybrid inflation model since, by construction, the
breaking of $G=U(1)$ at the end of inflation must generate cosmic
strings. This property of D-term inflation is totally independent
of the GUT framework or any other assumption.

\subsection{F-term Supersymmetric hybrid inflation}
As introduced in the previous section, F-term hybrid inflation
is based on the superpotential
\begin{equation}\label{ftermsuperpotential2}
W_{\rm F}=\kappa S (\Phi_+ \Phi_- - M^2)~,
\end{equation}
where $\kappa$ is the coupling constant between the inflaton and
the Higgs fields and $M$ is the mass scale parameter. $S$,
$\Phi_\pm$ are chiral superfields, and $S$ is a singlet under a
group $G$ while $\Phi_\pm$ belong to non trivial conjugate
representations of $G$. In the global SUSY framework, we can
derive the tree level scalar potential
\begin{equation}\label{Ftreelevel}
V_{\rm F}(\phi_+,\phi_-,S)=\kappa^2|M^2-\phi_-\phi_+|^2
+\kappa^2|S|^2(|\phi_-|^2+|\phi_+|^2)
+\mathrm{D-terms}~.
\end{equation}
At high inflaton values $S\gg S_c = M$ (we assume that chaotic
initial conditions for inflation will impose large values for the
inflaton field initially), the potential is minimized by the
configuration $\langle \phi_\pm\rangle =0$ ($\phi_\pm$ correspond
to the scalar components of the chiral superfields $\Phi_\pm$).
Thus we obtain a perfectly flat potential which is non vanishing
$V_0=\kappa^2 M^4$. As a result, SUSY is temporarily broken,
and this induces radiative corrections to the tree level
potential. The 1-loop contribution can be calculated by the
Coleman-Weinberg formulae leading to an effective inflation
potential \cite{jcap2005}
\begin{equation}\label{Feffective}
\begin{split}
V_{\rm eff}^{\rm F}(|S|)&=V_0+[\Delta V(|S|)]_{1-{\rm loop}}\\
&=\kappa^2 M^4 \biggl\{1+\frac{\kappa^2 \cal{N}}{32\pi^2}
\biggl[2\ln\frac{|S|^2\kappa^2}{\Lambda^2}+ (z+1)^2\ln(1+z^{-1})+
(z-1)^2\ln(1-z^{-1})\biggl]\biggl\}~,
\end{split}
\end{equation}
where $\mathcal{N}$ is the dimensionality of the representation of
$G$ to which $\Phi_\pm$ belongs and $\Lambda$ is a renormalization
scale. We have also defined $z \equiv S^2/S_c^2$. Note that, in the 
literature, papers often consider only the first term in the
radiative correction, corresponding to the limit $z\gg 1$. This
turns out to be not consistent since the last 60 e-folds of
inflation can occur for values of the inflaton field close to its
critical value $z=1$. What are typical values for $\mathcal{N}$ ?
Interestingly enough, this \emph{integer} is related to the choice
of GUT group such that  $G\subset\ggut$ and the SSB pattern. For
example, if we consider models based on $SO(10)$, the inflation
phase often occurs at the breaking of the $\bl$ symmetry. This can
be seen on SSB patterns through the Pati-Salam group where $\bl$
is readily identified [see Eq.~\ref{schemasso10}]. To protect the
R-parity subgroup of $\bl$, one needs a pair of $\mathbf{126}$ and
$\mathbf{\overline{126}}$ to break $\bl$. Thus, typically for $SO(10)$
GUT, $\mathcal{N}=126$. Typical values for $E_6$ Higgs
representations are $\mathbf{27}$ and $\mathbf{351}$.

The end of the inflation phase is triggered by the first of the
two following mechanisms :
\begin{itemize}
\item the Higgs fields become massive, and $\langle
\phi_+\rangle = \langle \phi_+\rangle =0$ is not stable anymore.
This condition is reached at $z=1$.

\item the end of the slow roll conditions : $\epsilon =\mathcal{O}(1)$
or $\eta =\mathcal{O}(1)$. We can show numerically \cite{jcap2005}
that this condition is also reached for $z\simeq 1$.
\end{itemize}
Once slow roll inflation ends, the dynamics will set the
system to the global minimum of the potential of
Eq.~(\ref{Ftreelevel}), namely $\langle S\rangle =0$, $\langle
\phi_\pm\rangle = M$. The symmetry $G$ under which $\Phi_\pm$ are
charged is broken. Note that the VEV that is responsible for
cosmic string formation is thus given by $M$.\\

The contribution from inflation to the CMB temperature anisotropies 
can be divided into two parts : the scalars and the tensors. They are
functions of the inflation potential and the value of the inflaton
fields at the considered scale. Hereafter we will focus on the
quadrupole anisotropy. At this scale, the contributions are given
by the Sachs-Wolfe effect and read
\begin{equation}\label{contribInflScal}
\left(\frac{\delta T}{T}\right)_{\rm Q- scal} =
\frac{1}{4\sqrt{45}\pi}\frac{V^{3/2}(S_Q)}{M_{\rm
Pl}^3\,V'(S_Q)}~,
\end{equation}
and
\begin{equation}
\label{contribInflTens}
\left(\frac{\delta T}{T}\right)_{\rm Q-tens}\sim {0.77\over 8\pi}
\,\frac{V^{1/2}(S_Q)}{M_{\rm Pl}^2}.
\end{equation}
where $\Mpl ={(8\pi G)^{-1/2}}\simeq 2.43\times 10^{18}$ GeV is the
reduced Planck mass and $S_Q$ is the value of the inflaton field
corresponding to the quadrupole. This value is another unknown as
well as the parameters $\kappa$ and $M$.

A first relation between these unknowns comes from the number of
e-folds between the quadrupole scale and the end of inflation.
This number reads
\begin{equation}
\label{nefold}
N_Q\equiv N(S_Q \rightarrow S_c)=-\frac{1}{\Mpl^2}
\int_{S_Q}^{S_c = M}{V(S) \over V'(S)}{\rm
d}S~.
\end{equation}
To solve the horizon problem, we require this number to be $N_Q
\simeq 60$. For a given value of $\mathcal{N}$ and $\kappa$, this
equation can give a relation between the inflaton value $S_Q$ and
the energy scale $M$. To completely fix this scale, we need to
normalize the total contribution to the COBE measurement. The
important point here to understand is that since cosmic strings
are formed at the end of inflation, we need to take their
contribution into account before normalizing. The contribution of
a local Nambu-Goto string network at the quadrupole scale is
proportional the mass per unit length $\mu$, $$\left(\frac{\delta
T}{T}\right)_{\rm cs}=\alpha G\mu~.$$ In first approximation, $\mu
= 2\pi \langle h \rangle ^2$ where $h$ is the VEV of the Higgs
field responsible for cosmic string formation. In our case,
$h=M$ and the numerical factor $\alpha$ is computed through heavy
numerical simulations. Recent results obtained in realistic
cosmologies give $\alpha=9-10$, and thus \cite{jcap2005}
\begin{equation}\label{contribCS}
\left(\frac{\delta T}{T}\right)_{\rm cs}\sim \frac{9}{4}
\frac{M^2}{\Mpl^2}~.
\end{equation}

\begin{figure}[hhh]
\includegraphics[scale=.6]{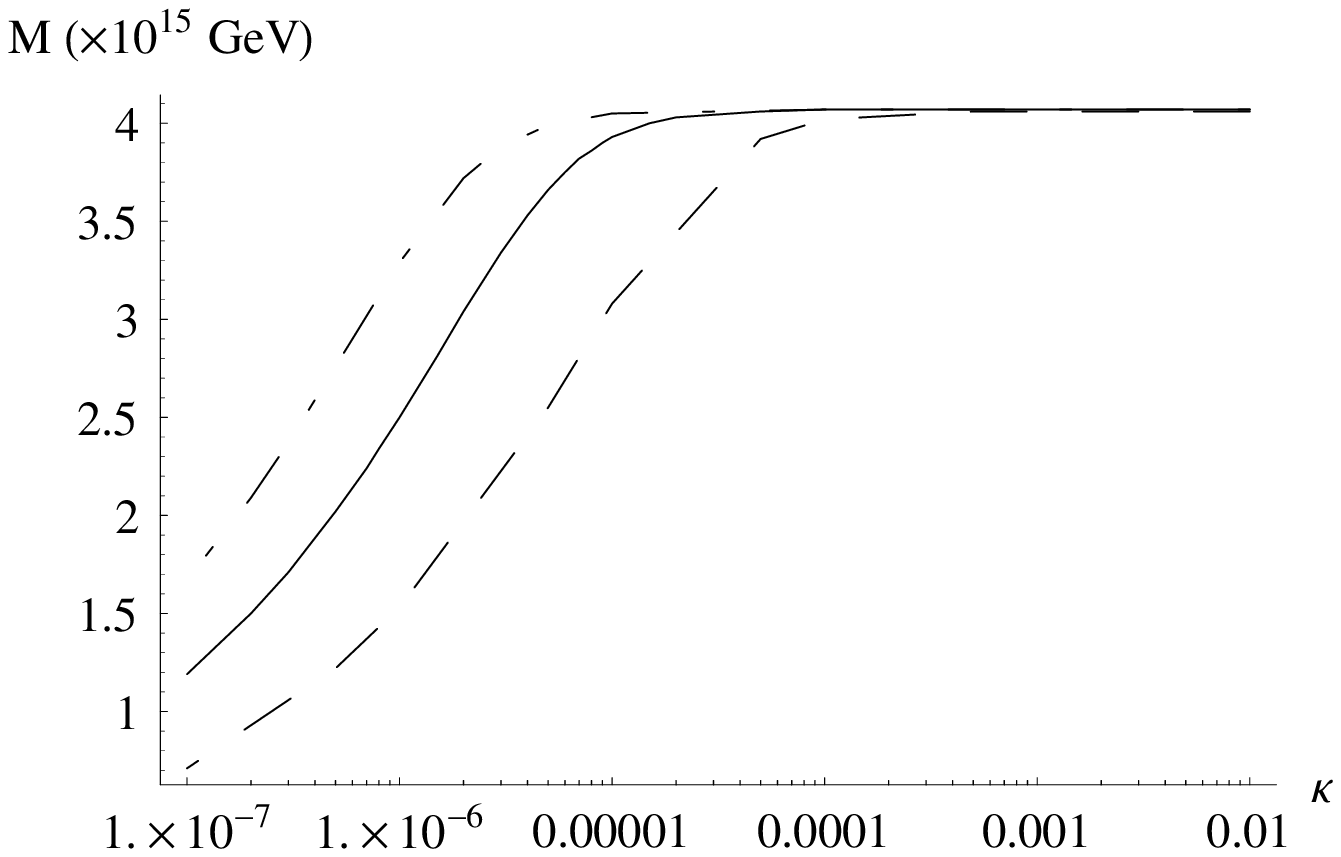}
\includegraphics[scale=.6]{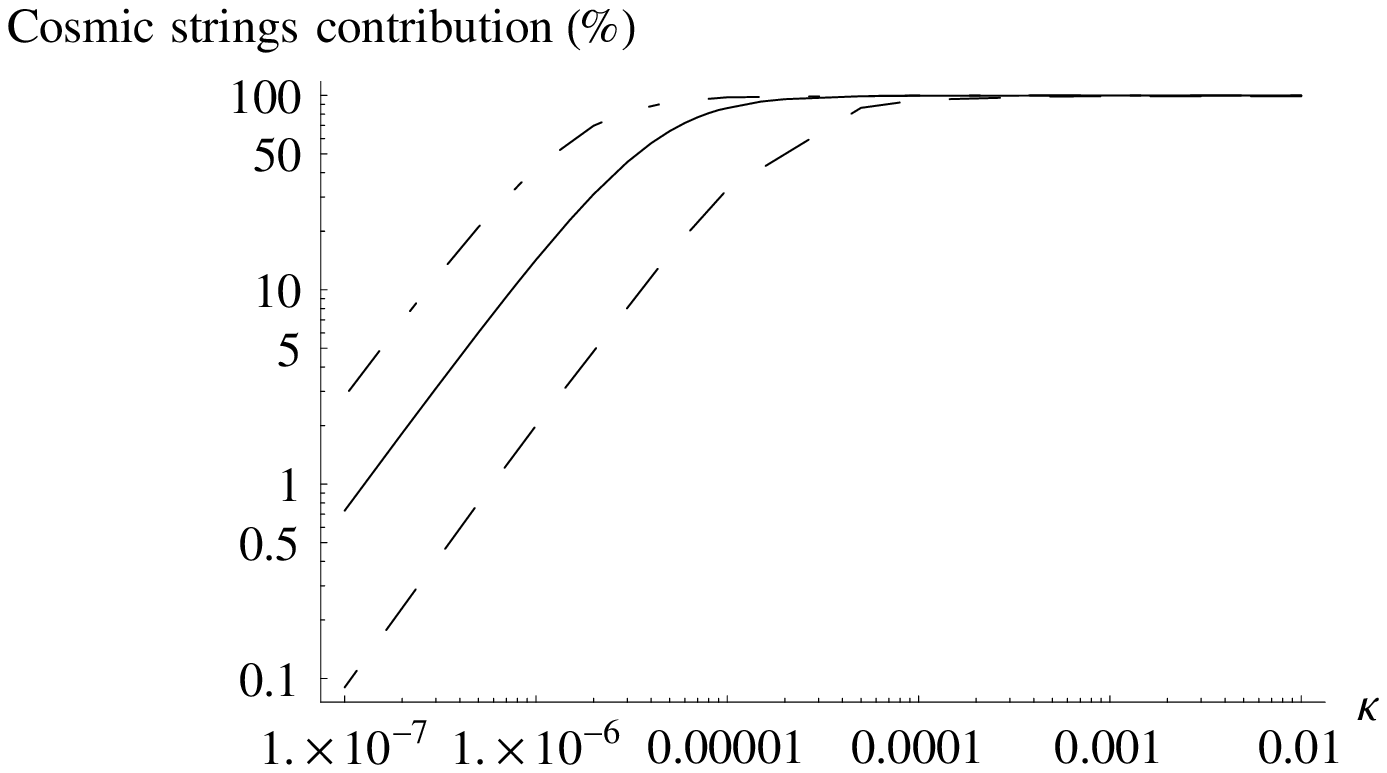}
\caption{On the left: Evolution of the inflationary scale $M$ in units
of $10^{15}$ GeV as a function of the dimensionless coupling $\kappa$.
On the right: Evolution of the cosmic strings contribution to the
quadrupole anisotropy as a function of the coupling  of the
superpotential, $\kappa$.
The three curves correspond to $\mathcal{N}=\mathbf{27}$ (curve with
broken line), $\mathcal{N}=\mathbf{126}$ (full line) and $\mathcal{N}
=\mathbf{351}$ (curve with lines and dots).}
\label{inflscaleF}
\end{figure}

Therefore, the total quadrupole anisotropy is given by
\begin{equation}\label{dttot}
\left[\left(\frac{\delta T}{T}\right)_{\rm Q-tot}\right]^2=
\left[\left(\frac{\delta T}{T}\right)_{\rm
scal}\right]^2+\left[\left(\frac{\delta T}{T}\right)_{\rm
tens}\right]^2+\left[\left(\frac{\delta T}{T}\right)_{\rm cs}\right]^2
\end{equation}
The normalization to the COBE measurement must then be done by
setting the left-handed side of this equation to $\left(\delta
T/ T\right)_{\rm Q}^{\rm COBE} \sim 6.3\times 10^{-6}$. This
allows us to calculate the energy scale $M(\kappa)$ as a function
of the coupling constant and for a given value of $\mathcal{N}$.
This is shown on Fig.~\ref{inflscaleF}.

We can also define the fraction of anisotropies generated by
cosmic strings
\begin{equation}
{\cal A}_{\rm cs}\sim\left[
{\left({\delta T\over T}\right)^{\rm cs}_{\rm Q}
\over
\left({\delta T\over T}\right)^{\rm COBE}_{\rm Q}}\right]^2~,
\end{equation}
which is also represented in Fig.~\ref{inflscaleF}.

A recent Bayesian analysis in a 6+2 dimensional parameter space
\cite{wyman} has shown that a cosmic string contribution to the
CMB fluctuations ${\cal A}_{\rm cs}$ higher than $14\%$ is
excluded by the WMAP 1st year data up to $99\%$ confidence level.
This can be translated into a contraint on $\kappa$ or $M$,
\begin{equation}\label{kconstr}
\kappa \lesssim 1\times 10^{-6} \times \frac{126}{\mathcal{N}}~,
\quad M \lesssim 2.5 \times 10^{15}\;\mathrm{GeV}~.
\end{equation}
These relations hold for $\mathcal{N}=1,16,27,126,351$. The first
conclusion \cite{jcap2005} is that contrary to what was previously
thought, F-term hybrid inflation suffer from fine tuning of its
superpotential coupling constant. We are also able to constrain
the inflation energy scale, the constraint being robust and
independent of the GUT group considered. Interestingly, this work
also shows a relation between a measurable quantity (the cosmic
string contribution to CMB) and a parameter that is directly
related to the fundamental GUT group of the model (the parameter
$\mathcal{N}$). Future improvement in CMB experiments and the
detection of a string contribution could bring new information on
high energy physics.

\subsection{D-term inflation}
Let's now turn to the D-term inflation model
\cite{binetdvali,binetAVP} and the constraints on the parameter
space imposed by the formation of cosmic strings. Here this
formation is not linked to the embedding of the inflation model
inside any framework. It is simply due to the fact that D-term
inflation is necessarily ended by the breaking of a $U(1)$ symmetry.

The procedure here is similar to the one of the previous
subsection for F-term inflation and we will just give here the
guidelines and the results for D-term inflation.

In this model, $\Phi_\pm$ must be charged under a U(1)
symmetry with, let's say, charges $\pm 1$. The inflaton
superfields $S$ remain trivially charged and the model requires
also a non-vanishing Fayet-Iliopoulos term $\xi$. The model is
finally based on a superpotential of the form
\begin{equation}
\label{superpotDbis}
W_{\rm D}=\lambda S \Phi_+\Phi_-~,
\end{equation}
where $\lambda$ denotes the superpotential coupling. We can show
\cite{prl2005,jcap2005} however that the global SUSY analysis for
this model is not consistent, since the inflaton field $S_Q$ can
reach plankian values even if the energy of the inflaton potential
stays below the Planck scale. This means that the analysis should
be performed in SUGRA. We will hereafter consider a minimal SUGRA,
i.e. the minimal structure for gauge kinetic function
$f_{ij}(\Phi_i)=\delta_{ij}$ and a K$\ddot{\rm a}$hler potential
given by
\begin{equation}\label{miniK}
K=|\phi_-|^2+|\phi_+|^2+|S|^2~.
\end{equation}
Therefore, as it was found in \cite{binetdvali}, the
scalar potential reads
\begin{eqnarray}\label{DpotenSUGRAtot}
V^{\rm D}_{\rm SUGRA} & =&
\lambda^2\exp\left({\frac{|\phi_-|^2+|\phi_+|^2+|S|^2}{M^2_{\rm
Pl}}}\right)
\biggl[|\phi_+\phi_-|^2\left(1+\frac{|S|^4}{M^4_{\rm
Pl}}\right)
\nonumber\\
&+& |\phi_+S|^2 \left(1+\frac{|\phi_-|^4}{M^4_{\rm
Pl}}\right)+|\phi_-S|^2 \left(1+\frac{|\phi_+|^4}{M^4_{\rm
Pl}}\right) +3\frac{|\phi_-\phi_+S|^2}{M^2_{\rm
Pl}}\biggl]\nonumber \\ &+&
\frac{g^2}{2}\left(|\phi_+|^2-|\phi_-|^2+\xi\right)^2~,
\end{eqnarray}
where $g$ is the gauge coupling of the $U(1)$ symmetry. Once
again, for large values of the inflaton field $S\gg S_c $ the
potential is minimized by $|\phi_+|=|\phi_-|=0$ and we get a
perfectly flat potential at tree-level $V_0=g^2\xi^2/2$. At one
loop, during the inflation phase, the effective scalar potential
is given by \cite{prl2005}
\begin{equation}
\label{VexactDsugra}
V^{\rm D-SUGRA}_{\rm eff} =
\frac{g^2\xi^2}{2}\left\{1+\frac{g^2}{16\pi^2}
\left[2\ln\frac{|S|^2\lambda^2}{\Lambda^2}\exp\left({|S|^2\over M_{\rm
Pl}^2}\right)+
(z+1)^2\ln(1+z^{-1})+(z-1)^2\ln(1-z^{-1})\right]\right\}~,
\end{equation}
where $z=[\lambda^2 |S|^2/(g^2\xi)]\exp(|S|^2/M_{\rm Pl}^2)$. We
can show that the end of inflation is reached when $z\simeq 1$ and
then the breaking of $U(1)$ and the formation of strings are
generated by $\langle\phi_+ \rangle=0$ and $\langle\phi_-\rangle
=\sqrt{\xi}$.

\begin{figure}[hhh]
\includegraphics[scale=.6]{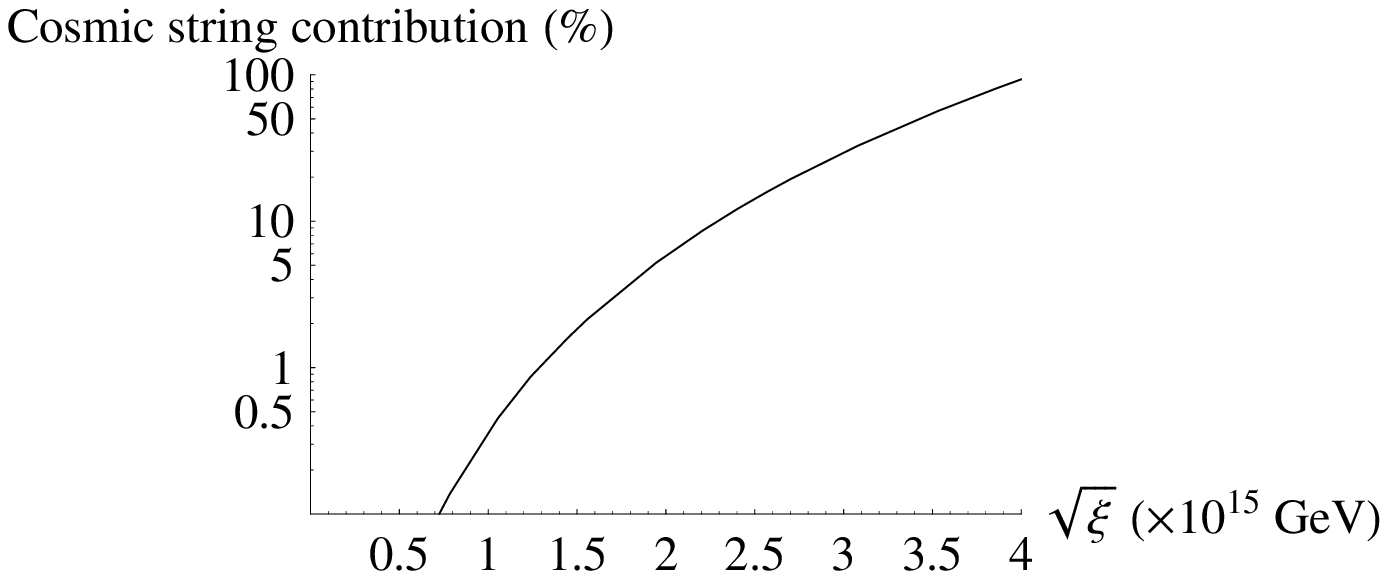}
\includegraphics[scale=.45]{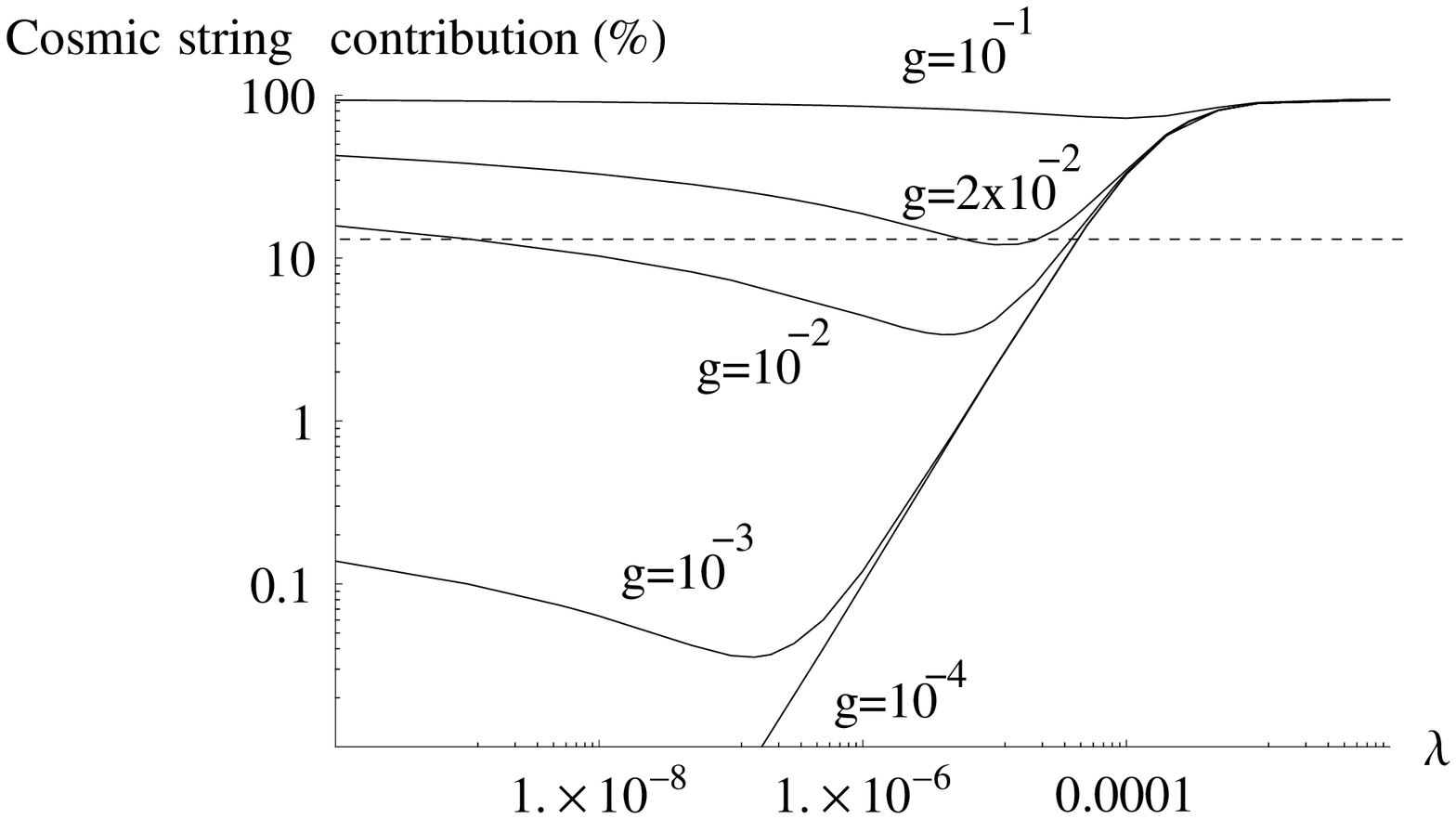}
\caption{On the left, the cosmic string contribution as a function of
the mass scale $\sqrt\xi$. This holds for all studied values of
$g$. On the right, the cosmic string contribution to the CMB
temperature anisotropies, as a function of the superpotential
coupling  $\lambda$, for different values of the gauge coupling
$g$. The maximal contribution allowed by WMAP is represented by a
dotted line. The plots are derived in the framework of SUGRA.}
\label{contribDsugra}
\end{figure}

The procedure to compute the mass scale $M\equiv\sqrt{\xi}$ as a
function of $\lambda$ and $g$ is similar to the case of F-term
inflation and the results are shown in Fig.~\ref{contribDsugra}.

The first conclusion is that D-term inflation can be in agreement
with CMB data even if cosmic strings of the GUT scale are formed
\cite{prl2005}. We can also see that all parameters of the model
are constrained by imposing an upper limit on the cosmic string
contribution to the CMB anisotropies. The mass scale possesses the
same limit as in the F-term case, at $99\%$ CL,
\begin{equation}
\sqrt\xi \lesssim 2.5\times 10^{15}\;{\rm GeV}~.
\end{equation}
which can be understood by the fact that this limit is equivalent to
the limit on the string contribution. The coupling constants of
the model are also constrained at $99\%$ of CL,
\begin{equation}
\lambda \lesssim 3\times 10^{-5}~,\quad g \lesssim 2\times 10^{-2}~.
\end{equation}
These results allow us to conclude that as for the F-term case,
D-term inflation can be in agreement with CMB data, but the price
is a fine tuning on its superpotential coupling constant. Note
that contrary to what has been previously found \cite{jcap2005,prl2005}, 
the inflaton field, in SUGRA, stays in the range $S_Q\in [10^{-3} - 10] 
\Mpl$ for typical 
values of $g$ and $\lambda$. Note also that D-term inflation has been
recently revisited \cite{binetAVP}. However, our results remain
valid since the modification of the model is of amplitude
$\xi/\Mpl^2 \lesssim 10^{-6}$.

\section{Conclusion}
In this paper, we have reviewed the formation of topological
defects within a large number of SUSY GUT models that are in
agreement with major observational constraints from particle
physics experiments and cosmology observations. Our goal is to get
a probability of cosmic string formation in this framework. The
conclusion is that cosmic strings are highly generic objects and
their energy is generically of the same order of magnitude as
the energy scale of inflation. As a consequence, they are
cosmologically relevant and we should continue searching for them
in various cosmological observations.

We analyzed in the second section the consequences of this
conclusion on two of the main inflationary models : F- and D-term
SUSY hybrid inflation. The generic formation of cosmic strings at
the end adds an additional contribution to the generation of CMB
anisotropies. This un-observed contribution is constrained by CMB
observations to be subdominant ($\lesssim 14\%$), and is employed
to constrain the parameters of the models. We have shown that they
can both be in agreement with most recent CMB data but face a fine
tuning of their superpotential coupling constant. We also show
that while the global SUSY description of F-term inflation can be
sufficient, this is not the case of D-term inflation, that must
be studied in SUGRA.

Several interesting questions suggested by these results remain
open : What is the generic micro-structure of cosmic strings
formed in our framework? How could this affect their impact on CMB
and the constraint on their contribution? In the case of F-term
inflation, is it possible to find a lower limit on the coupling
constant to really test the model with future experiments? Is our
conclusion for D-term inflation modified if we embed it in a non
minimal SUGRA description? What kind of SUGRA superpotentials and
K\"ahler potentials are motivated from high energy physics?


\begin{theacknowledgments}
It's a pleasure to thank the organizers of the Albert Einstein's Century
Meeting for inviting me to give this talk.
\end{theacknowledgments}



\bibliographystyle{aipproc}   

\bibliography{bibproc_Rocher}

\IfFileExists{\jobname.bbl}{}
 {\typeout{}
  \typeout{******************************************}
  \typeout{** Please run "bibtex \jobname" to optain}
  \typeout{** the bibliography and then re-run LaTeX}
  \typeout{** twice to fix the references!}
  \typeout{******************************************}
  \typeout{}
 }

\end{document}